\documentclass[onecollarge,natbib]{svjour2}
\bibpunct{[}{]}{;}{n}{}{,} 
\smartqed  
\usepackage{graphicx}

\usepackage{mathptmx}      
\journalname{Proceedings of Light Cone 2013}
\begin{document}

\title{Valence double parton distributions of the nucleon in a simple model
\thanks{Talk presended by WB at the Light Cone 2013 Conference, Skiathos, Greece. 
Supported by Polish Ministry of Science and Higher Education (grant DEC-2011/01/B/ST2/03915),  
and by Spanish DGI (grant FIS2011-24149) and Junta de Andaluc{\'{\i}a} (grant FQM225).         }
}


\author{Wojciech Broniowski         \and
        Enrique Ruiz Arriola 
}


\institute{Wojciech Broniowski \at
              The H. Niewodnicza\'nski Institute of Nuclear Physics, polish Academy of Sciences, PL-31342 Krak\'ow, Poland, and \\ 
              Jan Kochanowski University, PL-25406 Kielce, Poland \\
              \email{Wojciech.Broniowski@ifj.edu.pl}           
           \and
           Enrique Ruiz Arriola \at
              Departamento de F\'{\i}sica At\'{o}mica, Molecular y Nuclear and Instituto Carlos I de  F{\'\i}sica Te\'orica y Computacional. 
              Universidad de Granada, E-18071 Granada, Spain
}

\date{31 October 2013}

\maketitle

\begin{abstract}
Valence double parton distribution functions of the nucleon are
evaluated in the framework of a simple model, where the conservation
of the longitudinal momentum is taken into account.  The leading-order
DGLAP QCD evolution from the low quark-model scale to higher
renormalization scales is carried out via the Mellin moments of the
distributions. Results of the valence quark correlation function show
that in general the double distributions cannot be
approximated as a product of the single-particle distributions.
\end{abstract}

\vspace{1cm}

Theoretical interest in double parton distributions (dPDFs) in hadrons has
been recently renewed (see the
reviews~\cite{Bartalini:2011jp,Snigirev:2011zz} and the references
therein). The activity is largely triggered be the experimental
program at the LHC, where the double parton scattering (DPS) may bring
substantial contribution in certain production processes~\cite{d'Enterria:2012qx,Luszczak:2011zp}.  
Phenomenological aspects of the inclusion of the multi-parton effects 
have been widely pursued since the early days of
the parton model~\cite{Kuti:1971ph},
while many formal features of the multiparton evolution~\cite{Kirschner:1979im,Shelest:1982dg} follow from a pioneering
study on fragmentation functions~\cite{Konishi:1979cb}.
Elements of the QCD formulation are presented in~\cite{Collins:1981uk,Diehl:2010dr,Diehl:2011tt,Diehl:2011yj},
while a  ${}_2$GPD interpretation is provided in~\cite{Blok:2010ge}.
Positivity bounds for dPDFs were recently discussed in~\cite{Diehl:2013mla}.

Dynamic modeling of dPDFs has been up to now very little explored.  In
fact, the only calculations which nonperturbatively predict the
valence dPDFs of the nucleon has been carried out in the framework of
the MIT bag model~\cite{Chang:2012nw} and in the constituent quark
model~\cite{Rinaldi:2013vpa}.
  In both calculations, the approximate implementation
  of relativity has the side effect of producing distributions with
  unphysical support, i.e., extending outside the interval $0 \le x_1, x_2$, $x_1+x_2 \le 1$.  
  An important aspect of the dPDFs is that they allow to detect
  tight diquark correlations, in which case the dPDF becomes a function of the sum
  $x_1+x_2$. In the $(x_1,x_2)$ distribution plot the fingerprints are the
  straight-line structures inclined by $45^o$.

The purpose of this talk is to explore the valence dPDF of the proton
in a very simple model which includes the exact momentum-conservation
constraints and complies to the Lorentz invariance.
Satisfaction of these requirements results in proper theoretical
features, such as the correct support in the $x$-variables or the
quark-number and momentum sum rules~\cite{Gaunt:2009re}. In the
considered model, the correlations between partons are due solely to
the longitudinal momentum conservation. Importantly, we carry out
the LO DGLAP evolution~\cite{Kirschner:1979im,Shelest:1982dg} of the
obtained valence dPDF, confirming its crucial effects for the
corresponding quark correlation function as the evolution scale is moved up to
the experimentally-accessible values. We find that with the increasing
renormalization scale, the correlation between the valence quarks becomes
large and positive along the edges where $x_1$ or $x_2$ is small,
while it is negative in the region where $x_1$ and $x_2$ are similar and
relatively large. This shows in explicit terms that approximating dPDFs
with a product of two single-parton distributions (sPDF) is not
justified.

Similarly to sPDF, which is interpreted as the probability
distribution for the probed parton to carry the fraction $x$ of the
total $p^+$ momentum of the hadron, the dPDF has the interpretation of the joint
probabilistic distribution that the $j_1$ and $j_2$ partons carry the
momentum fractions $x_1$ and $x_2$ of the hadron, respectively. Denoting the sPDF as
$D_j(x)$ and dPDF as $D_{j_1 j_2}(x_!,x_2)$, one may define the parton correlation
function~\cite{Chang:2012nw,Rinaldi:2013vpa}
\begin{eqnarray}
\rho_{j_1 j_2}(x_1, x_2)=\frac{D_{j_1 j_2}(x_1, x_2)}{D_{j_1}(x_1) D_{j_2}(x_2)}-1. \label{eq:rho}
\end{eqnarray}

As mentioned, a dynamical nonperturbative calculation for the proton
has been made in the MIT bag model~\cite{Chang:2012nw}, where the
correlations stem from the conservation of the linear momentum
(implemented in terms of the Peierls-Yoccoz projection).  As is
well known, projecting onto good linear momentum coincides with
boosting from the rest frame only for the exact eigenstates of
relativistic systems~\cite{Betz:1983dy}. A constituent quark model
study has been carried out in~\cite{Rinaldi:2013vpa}.  Here we
consider a much simpler model which in our view grasps the essential
features of the problem and has the proper support.
The role of transverse degrees of freedom is discussed at the end. 

Let us introduce the probability that the parton carries the momentum
fraction $x$ as the square of its wave function,
$\phi(x)=|\psi(x)|^2$. Then the {\em three-particle} probability
distribution is proposed in the form
\begin{eqnarray}
\hspace{-7mm} D_3(x_1,x_2,x_3) = \phi(x_1)\phi(x_2)\phi(x_3)\delta(1-x_1-x_2-x_3), \label{eq:D3}
\end{eqnarray}
where the delta function enforces the longitudinal momentum conservation.
Its marginal projections define the dPDF and sPDF:
\begin{eqnarray}
&& \hspace{-11mm} D_2(x_1,x_2) = \int_0^1 dx_3 D_3(x_1,x_2,x_3) = \phi(x_1)\phi(x_2)\phi(1-x_1-x_2), \label{eq:D2} \\
&& \hspace{-11mm} D_1(x_1) = \int_0^1 dx_2 D_2(x_1,x_2)= \int_0^1 \theta(1-x_1-x_2) dx_2 \phi(x_1)\phi(x_2)\phi(1-x_1-x_2). \label{eq:D1} 
\end{eqnarray}
Note that Eq.~(\ref{eq:D1}) is equivalent to the quark number sum
rule of Gaunt and Stirling (GS)~\cite{Gaunt:2009re}.  The momentum GS sum
rule is also satisfied, as it is straightforward to show that
\begin{eqnarray}
&& \hspace{-11mm} \int_0^{1-x_1} dx_2\,x_2 D_2(x_1,x_2)+\int_0^{1-x_1} dx_3\,x_3 D_2(x_1,x_3)= (1-x_1)D_1(x_1). \label{eq:GSmom}
\end{eqnarray}
As a matter of fact, the only feature used in these derivations is the symmetric form of Eq.~(\ref{eq:D3}). 
{In addition, due to the symmetry, the extremum 
of $D_3(x_1,x_2,x_3)$ is located at the symmetric point $x_1=x_2=x_3=1/3$ regardless 
of its particular shape.} 

Early multiparton models~\cite{Kuti:1971ph} take the simple Regge-motivated parametrization $\psi(x) \sim x^a$ with \mbox{$a=1-\alpha(0)$} and
the intercept $\alpha(0)=1/2$, while the constant phase-space model of
Ref.~\cite{RuizArriola:1999hk} uses $\psi(x)=1$.  In this presentation
we investigate two simple classes of models of the form
\begin{eqnarray}
\phi(x)\equiv |\psi(x)|^2 = A x^\alpha \;{\rm (model~I)}, \;\;\;\; \phi(x)\equiv |\psi(x)|^2 = A (1-x)^a \;{\rm (model~II)}, \label{eq:phi} 
\end{eqnarray}
where $A$ is a normalization constant such that $\int_0^1 dx \, \phi(x)=1$. 
We consider the cases $\alpha=0,1/2,1$ and $a=1,2$. The first class
corresponds to the so-called {\em
  valon} model~\cite{Hwa:1980mv,Hwa:2002mv} for the proton, which
assumes $G_{uud/p} (x_1,x_2,x_3)= (x_1 x_2)^\alpha x_3^\beta \delta
(1-x_1-x_2-x_3)$ for $\alpha=\beta$.  For instance, for the case of the second class with $a=2$ in Eq.~(\ref{eq:phi}) we get 
\begin{eqnarray}
D_2(x_1,x_2) &=& \frac{1008}{29}(1-x_1)^2 (1-x_2)^2(x_1+x_2)^2, \nonumber \\
D_1(x)&=&\frac{168}{145}(1-x)^3(1+6x+16x^2+6x^3+x^4)
\end{eqnarray}
We note that the behavior of $D_1(x)$
near $x \to 1$ conforms to the QCD counting
rules~\cite{Lepage:1980fj}. This is also the case for $\alpha=1$ in
model~I of Eq.~(\ref{eq:phi}). As we see, at small $x$'s we can trace a diquark
correlation.

The above forms hold by construction at the {\em quark-model scale},
where the only degrees of freedom are the valence quarks.  To relate
to results at experimental scales, the appropriate QCD evolution is
necessary~\cite{Broniowski:2008tg}, which generates radiatively the
sea quarks and gluons.  The matching prescription based on the
requirement that at $\mu=2$~GeV the valence quarks carry the
experimental momentum fraction of 41.6\%~\cite{Ball:2012cx} leads to a
very low quark model scale, $\mu_0=285$~MeV (where the valence quarks
carry all the momentum). This low value is compatible to the case of
the pion in chiral quark models~\cite{Davidson:1994uv,Broniowski:2007si}.

The QCD evolution equations for multi-parton distributions have been derived long ago~\cite{Kirschner:1979im,Shelest:1982dg}. A simple and 
practical method of solving them numerically is based on the Mellin moments, similarly to the case of sPDFs. 
One introduces the corresponding moments of the sPDFs and dPDFs, 
\begin{eqnarray}
&& \hspace{-13mm} M^{n}_{j}=\int_0^1 dx \, x^n  D_j(x), \\
&& \hspace{-13mm} M^{n_1 n_2}_{j_1 j_2}=\int_0^1 \!\!dx_1 \int_0^1\!\! dx_2 \theta(1\!-\!x_1\!-\!x_2) x_1^{n_1} x_2^{n_2} D_{j_1 j_2}(x_1,x_2),  \nonumber
\end{eqnarray}
and the moments of the QCD splitting functions
\begin{eqnarray}
&& P_{i \to j}^n=\int_0^1 dx \, x^n P_{i \to j}(x), \\
&& P_{i \to j_1 j_2}^{n_1 n_2}=\int_0^1 dx \, x^{n_1} (1-x)^{n_2} P_{i \to j_1 j_2}(x), \nonumber \\
&& {\tilde{P}}_{i \to j_1 j_2}^{n_1 n_2}=\delta_{j_1 j_2}P_{i \to j_1}^{n_1+n_2} - 
 \delta_{i j_1}P_{j_1 \to j_2}^{n_2}  - \delta_{i j_2}P_{j_2 \to j_1}^{n_1}. \nonumber
\end{eqnarray}
Then the LO dDGLAP evolution equations read~\cite{Kirschner:1979im,Shelest:1982dg}
\begin{eqnarray}
&& \frac{d}{dt} M^{n_1 n_2}_{j_1 j_2} = \sum_i P_{i \to j_1}^{n_1} M^{n_1 n_2}_{i j_2} + \sum_i P_{i \to j_2}^{n_2} M^{n_1 n_2}_{j_1 i} 
+ \sum_i \left ( P_{i \to j_1 j_2}^{n_1 n_2} + {\tilde{P}}_{i \to j_1 j_2}^{n_1 n_2} \right) 
M^{n_1+n_2}_{i}, \label{eq:dDGLAP}
\end{eqnarray}
where the evolution variable is
\begin{eqnarray}
&& t= \frac{1}{2\pi \beta} \log \left [ 1+\alpha_s(\mu) \beta \log(\Lambda_{\rm QCD}/\mu) \right ], \;\; \beta=\frac{11N_c-2N_f}{12\pi}. \nonumber
\end{eqnarray}
Partons $i$, $j_1$, and $j_2$ may in general represent the valence quarks, the sea quarks, or the gluons, whose 
distributions are coupled. Moreover, the last term in Eq.~(\ref{eq:dDGLAP}) couples dPDFs to sPDFs.
In this talk we consider only the evolution of the valence quarks which leads to a technical simplification, as in this 
case there are no partons $i$ decaying into a pair of valence quarks and $P_{i \to j_1 j_2}=0$. This means that the 
inhomogeneous term in Eq.~({\ref{eq:dDGLAP}) vanishes and our evolution equation simply reads
\begin{eqnarray}
\frac{d}{dt} M^{n_1 n_2}_{j_1 j_2}(t) = \left ( P_{j_1 \to j_1}^{n_1} + P_{j_2 \to j_2}^{n_2} \right ) M^{n_1 n_2}_{j_1, j_2}(t).  \label{eq:evolut}
\end{eqnarray}
Recall that for sPDFs the corresponding evolution equation is 
\begin{eqnarray}
\frac{d}{dt} M^{n}_{j}(t) = P_{j \to j}^{n} M^{n}_{j}(t).  \label{eq:evoluts}
\end{eqnarray}
\begin{figure}[tb]
\begin{center}
\includegraphics[angle=0,width=0.4 \textwidth]{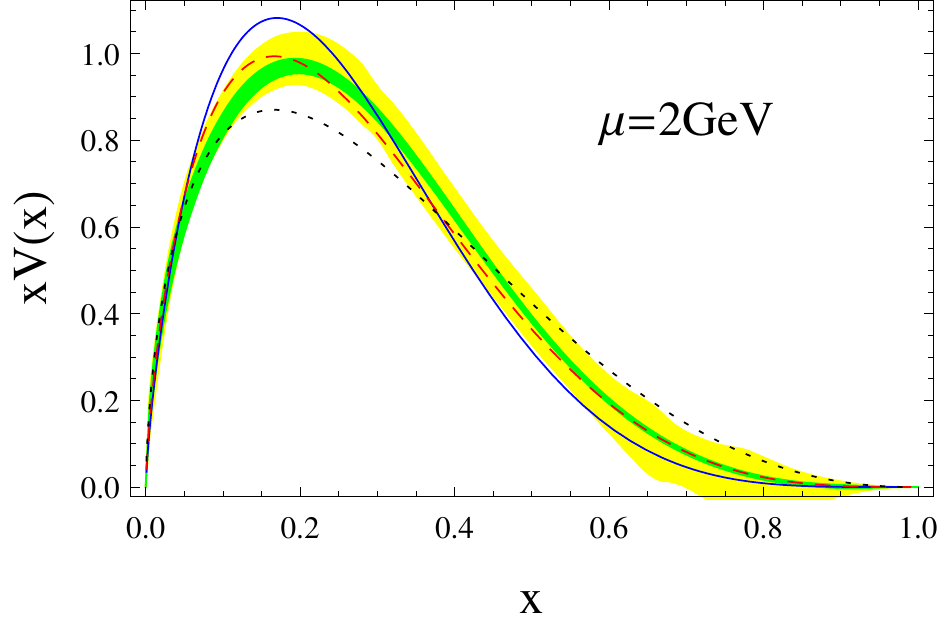} 
\includegraphics[angle=0,width=0.4 \textwidth]{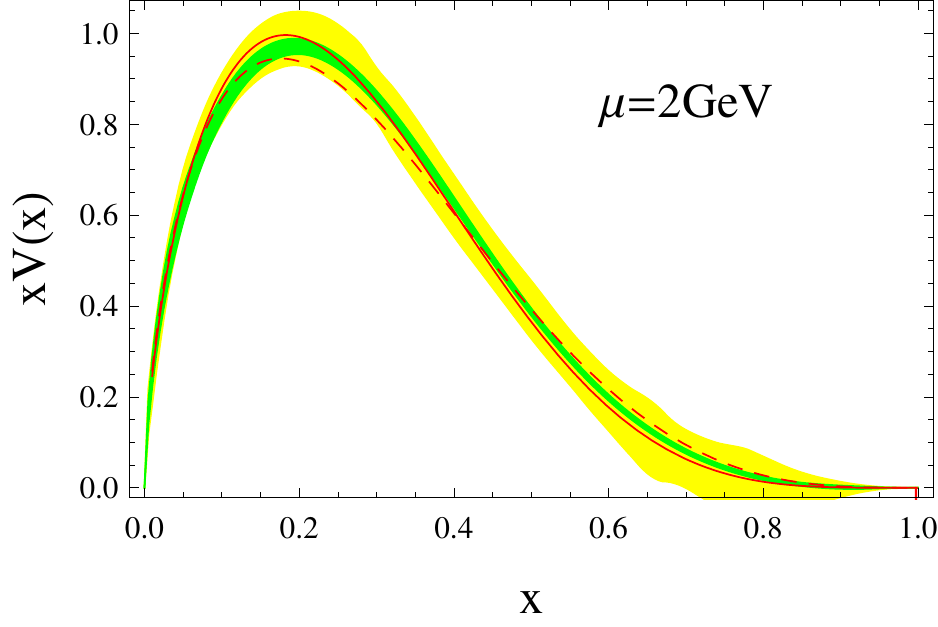} 
\end{center}
\caption{The valence sPDF of the nucleon (multiplied by $x$) at the
  scale of $\mu=2$~GeV, plotted as a function of the Bjorken variable
  $x$.  The darker band corresponds to the NNPDF2.3 fit with no LHC
  data, while the broader light band is the the NNPDF2.3 fit with the
  collider data only~\cite{Ball:2012cx}. We show sPDF evolved with the dDGLAP
  equations from the quark-model scale $\mu_0$ to $\mu=2$~GeV.
 {Left: } The solid, dashed, and dotted
  lines correspond to the valon model $|\psi(x)|^2=A x^\alpha$ with
  $\alpha=1$, $1/2$, and $0$, respectively. {Right: }  
The solid and dashed lines correspond to the model $|\psi(x)|^2=A (1-x)^a$ with $a=2$ and $a=1$,
respectively.}
\label{fig:sPDF}
\end{figure}  
Note that due to the presence of correlations we have $M^{n_1 n_2}_{j_1 j_2}(t)\neq M^{n_1}_{j_1}(t)M^{n_2}_{j_2}(t)$ and the system 
is not separable. The solutions of Eq.~(\ref{eq:evoluts},\ref{eq:evolut}) are
\begin{eqnarray}
&& M^{n}_{j}(t) = e^{P_{j \to j}^{n}(t-t_0)} M^{n}_{j}(t_0), \label{eq:smomevol} \\
&& M^{n_1 n_2}_{j_1 j_2}(t) = e^{(P_{j_1 \to j_1}^{n_1} + P_{j_2 \to j_2}^{n_2})(t-t_0)} M^{n_1 n_2}_{j_1 j_2}(t_0). \label{eq:dmomevol}
\end{eqnarray}

The inverse Mellin transform brings us to the evolved solution in the $x$-space, namely
\begin{eqnarray}
&& \hspace{-12mm} D_j(x;t) = \int_C \frac{dn}{2\pi i} x^{-n-1} M^{n}_{j}(t), \label{eq:mel1} \\
&& \hspace{-12mm} D_{j_1 j_2}(x_1,x_2;t) = \int_C \!\! \frac{dn_1}{2\pi i} x_1^{-n_1-1}  
\int_{C'} \!\! \frac{dn_2}{2\pi i} x_2^{-n_2-1}M^{n_1,n_2}_{j_1,j_2}(t), \nonumber \\ \label{eq:mel2}
\end{eqnarray}
where $n$ and $n'$ are treated as complex variables and the contours
$C$ and $C'$, lying right to all singularities of $M$, are
bended by $45^o$ (see e.g. Ref.~\cite{RuizArriola:1998er}).
The bending of the contours helps with the pace of
numerical convergence. Schwarz's reflection principle $M_{n^*}= M_n^*$
is satisfied since the anomalous dimensions are real for any integer
$n$ and thus one segment of the bended contour is needed in the evaluation.
The method is most practical when the moments $M$ are analytic
functions of their arguments; if they were not (as could be the case
when some numerical fits to the data are made), then they should be
approximated with analytic functions, e.g., with sums of the Euler
Beta functions, whence the procedure may be carried out with no
difficulty.

Next, we proceed to presenting our results of the evolution in the
considered model.  Before showing the main findings for the valence
dPDF, we explore the valence sPDF, where the results can be
straightforwardly compared to the existing data
parameterizations. Figure~\ref{fig:sPDF} presents the valence quark
distribution, customarily multiplied by $x$, evolved to the benchmark
scale of $\mu=2$~GeV and compared to the NNPDF2.3
fits~\cite{Ball:2012cx}, indicated with bands (the darker narrow band is
the fit excluding the LHC data, and the broad band uses the collider
data only).
\begin{figure*}[tb]
\begin{center}
~\hspace{-6.5mm}\includegraphics[angle=0,width=0.345
    \textwidth]{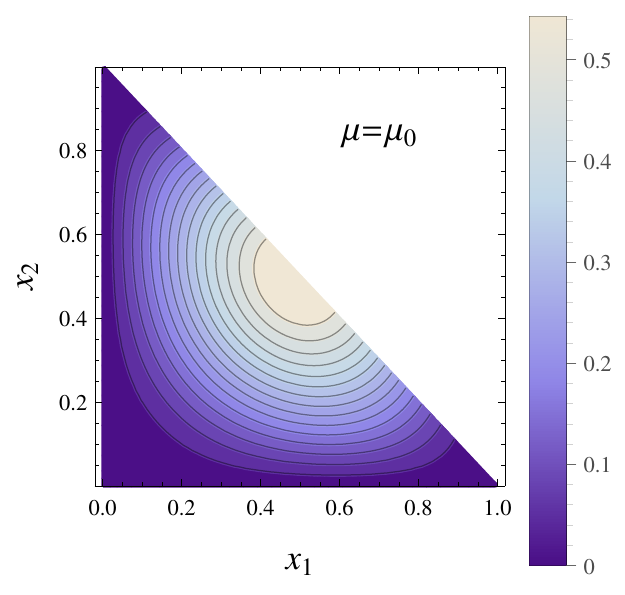}\includegraphics[angle=0,width=0.345
    \textwidth]{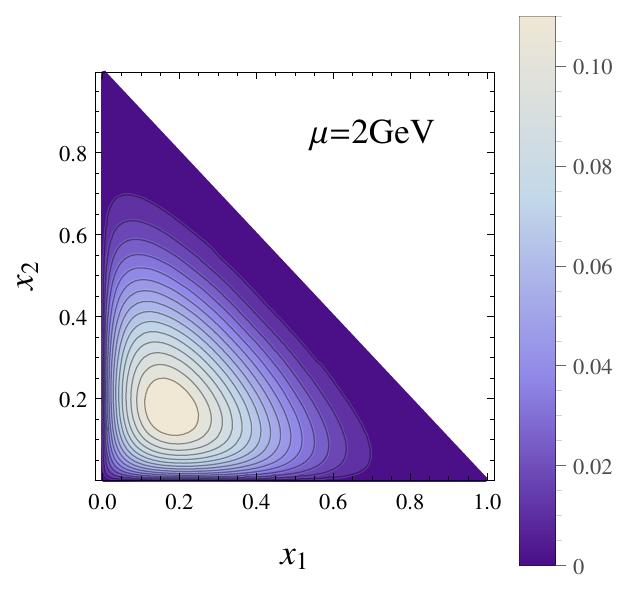}\includegraphics[angle=0,width=0.345
    \textwidth]{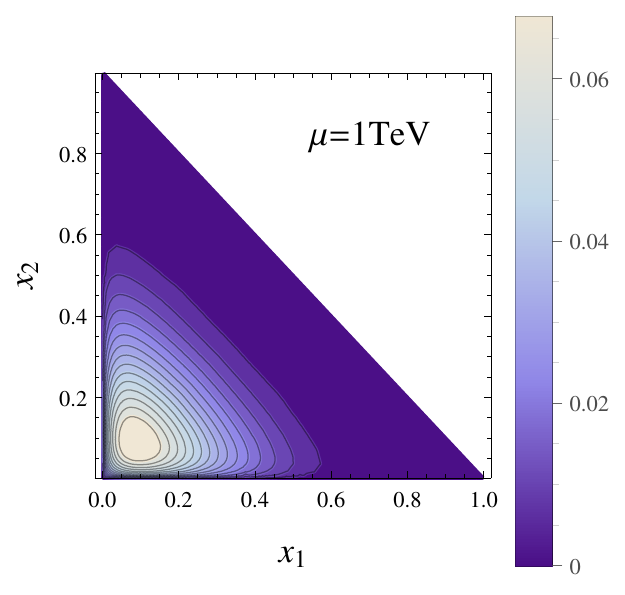}
\end{center}
\vspace{-2mm}
\caption{Contour plots of the valence dPDF of the nucleon multiplied
  by $x_1 x_2$, i.e., the quantity $x_1 x_2 D_2(x_1,x_2)$, at the scale
  $\mu_0$ and evolved to $\mu=2$~GeV and $\mu=1$~TeV with the LO
  dDGLAP equations. Model with $a=2$ corresponding to 
  $\psi(x)=\sqrt{3}(1-x)$.
\label{fig:dPDF}} 
\end{figure*}  
The reasonable reproduction of the valence sPDF gives us confidence
that the simple model of Eq.~(\ref{eq:D2},\ref{eq:D1}) grasps the most
essential features of the valence parton distributions of the nucleon,
particularly for the case $a=2$, $\psi(x)=\sqrt{3}(1-x)$. We may thus
move towards the double distributions.  The effect of the evolution on
the valence dPDFs is presented in Fig.~\ref{fig:dPDF}, where we show
the contour maps of $x_1 x_2 D_2(x_1,x_2)$ at three scales: the
quark-model scale $\mu_0$, and the higher scales $\mu=2$~GeV and $\mu=1$~TeV. Of course,
the support of the function is the region $x_1+x_2 \leq 1$, $x_i \geq
0$.  We observe how the dDGLAP equations cause the drift of the
distributions towards low values of $x_1$ and $x_2$. This drift to low
values of $x$ is well-known for the case of the DGLAP evolution of
sPDFs; it has also been observed for the dPDFs~\cite{Gaunt:2009re}. We
note the lack of factorization in the $x_1$ and $x_2$ variables within
the DGLAP approach, as emphasized in
Ref.~\cite{Snigirev:2003cq,Korotkikh:2004bz}.
\begin{figure*}[tb]
\begin{center}
~\hspace{-6.5mm}\includegraphics[angle=0,width=0.345\textwidth]{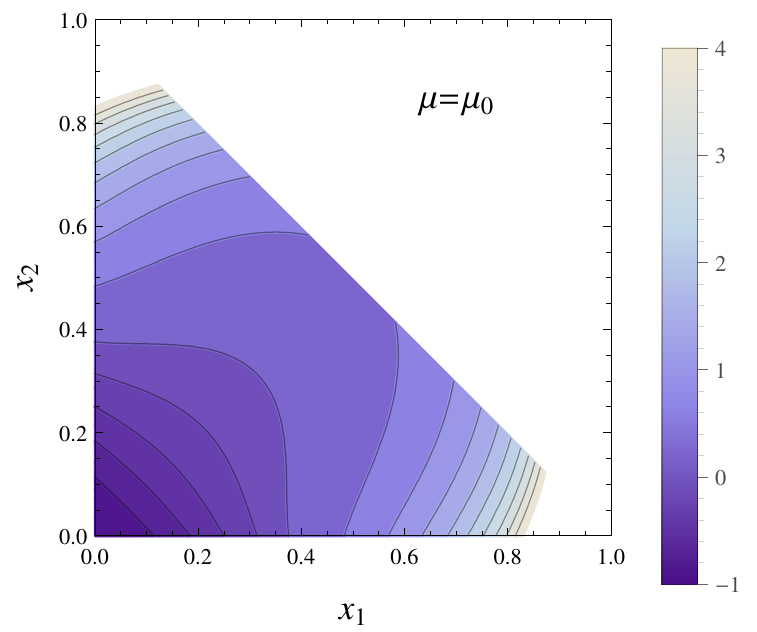}\includegraphics[angle=0,
width=0.345\textwidth]{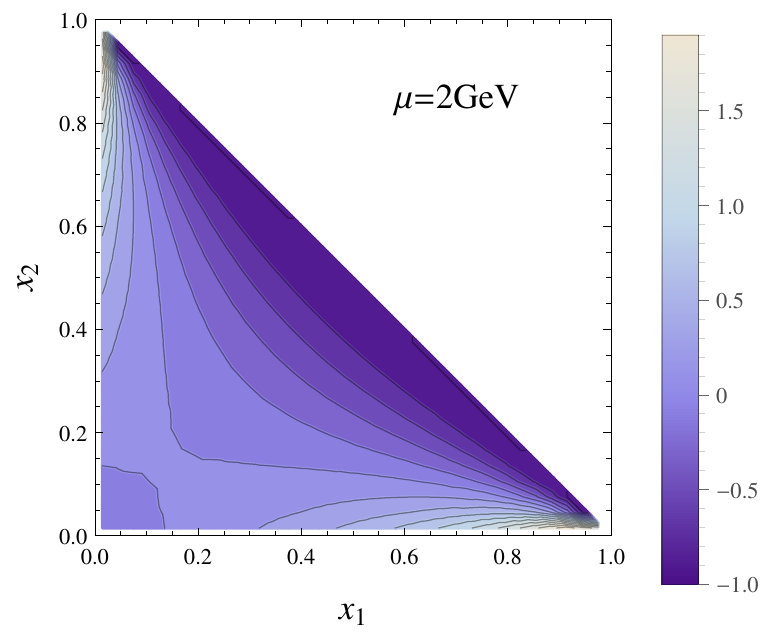}\includegraphics[angle=0,width=0.345\textwidth]{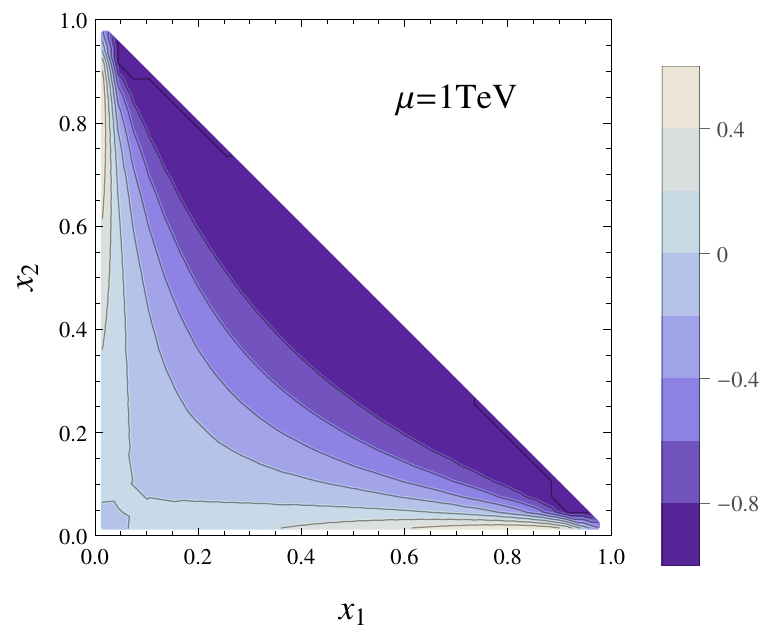} 
\end{center}
\vspace{-2mm}
\caption{Contour plots of the valence quark correlation function $\rho(x_1,x_2)=D_2(x_1,x_2)/[D_1(x_1)D_1(x_2)]-1$, 
at the scales $\mu_0$, $\mu=2$~GeV, and  $\mu=1$~TeV. Model with $a=2$ corresponding to 
  $\psi(x)=\sqrt{3}(1-x)$. 
\label{fig:cor}} 
\end{figure*}  
The behavior at the end-lines~\cite{Snigirev:2010ds} shows that for a
fast variable and a slow variable, i.e., $x_1 \to 1 $ and $x_2 \to 0$,
the correlation becomes small, since
\begin{eqnarray}
D_2(x_1,x_2;t) \to H(t) (1-x_1)^{k_{i,j_1,j_2}}  (1-x_1-x_2)^{2 C_F t + h_{i,j_1,j_2}}, 
\end{eqnarray}
a feature confirmed with our numerical analysis.

For a probability interpretation to hold, it is necessary that the evolution
equations preserve the positivity of the initial condition. Amusingly, it
has been found that the upward evolution for sPDFs produces positive
distributions~\cite{LlewellynSmith:1978me}, unlike the downward evolution
where a lower bound for the evolution ratio is found on this
basis~\cite{RuizArriola:1998er} (cf. also the  irreversible evolution
of Ref.~\cite{Teryaev:2005uf}).  
Within the considered model, the upward evolution preserves the
positivity of the valence dPDF, as can be seen from Fig.~\ref{fig:dPDF}.

In Fig.~\ref{fig:cor} we plot the double-parton correlation function
as defined in Eq.~(\ref{eq:rho}). As the scale increases, the
correlation becomes large and positive along the lines $x_i=0$, while
in the central region it is negative. At the line of the kinematic
constraint $x_1+x_2=1$ we find $\rho(x_1,x_2)=-1$.  The results of
Fig.~\ref{fig:cor} show that the valence dPDF cannot be approximated
with the uncorrelated product of the sPDFs.  We note that our plots
bare qualitative similarity to the results obtained in the MIT bag
model~\cite{Chang:2012nw}. 

We end with some remarks concerning the longitudinal and transverse degrees of freedom.
Our calculation is essentially one-dimensional, as the transverse
dynamics plays no active role. This dimensional reduction could be
possible if the relative momentum between the two quarks is assumed to
be small.  To analyze this assumption from a non-perturbative point of
view, let us remind ourselves that the transverse lattice
approach~\cite{Burkardt:2001jg} allows to effectively freeze the
transverse degrees of freedom when the transverse spacing, $\vec
a_\perp $, becomes larger than the characteristic resolution of the
parton momentum in the hadron, $\sim 1 /\Lambda_{\rm QCD}$. This also
limits the multiparton Fock state by powers of $a_\perp$. For
instance, for the nucleon $N = q^3 + a_\perp^3 q^4 \bar q + a_\perp^4 q^3 G+
\dots$. However, as soon as $a_\perp$ starts decreasing, gluons and
sea quarks are radiated and it is amazing that the DGLAP evolution
equations roughly reproduce the observed behavior for sPDF~\cite{Broniowski:2007si}.

A simple way to mimic the transverse lattice dynamics is by using a
2-dimensional harmonic oscillator to trap the partons in the
hadron~\cite{Vary:2009gt} in the transverse plane, with a mass squared
operator given by
\begin{eqnarray}
M^2 = \sum_i \left[ \frac{\vec k_{i\perp}^2+m_i^2}{x_i}+ \sigma ^2 x_i \vec b_{i}^2 \right] .
\label{eq:M2}
\end{eqnarray} 
Here, $x_i$ are the longitudinal momentum fractions fulfilling $\sum_i
x_i=1$, $\vec b_i$ are the impact-parameter variables,  $\vec
k_{\perp,i}$ their corresponding Fourier-conjugate variables, and $\sigma$ is
the oscillator constant. If $\psi_{n_\perp} ( x, \vec b) $ is the
single particle two dimensional HO wave function normalized as $\int
d^2 b |\psi_{n_\perp} ( x, \vec b)|^2 = 1$ with quantum numbers $\vec
n_\perp=(n_x,n_y)$ and mass squared $M^2 = 2 \sigma ( n_x+n_y + 1)$,
then the general solution is a symmetrized product of  factorized states 
\begin{eqnarray}
\Psi_{\vec n_{\perp}} (x_1,\vec b_1, \dots , x_N ,\vec b_N) = 
\varphi(x_1 , \dots , x_N) \prod_{i=1}^N \psi_{n_{i,\perp}} ( x_i, \vec b_{i}) \chi_{\rm spin-flavor}.
\end{eqnarray}
Due to the fact that the single-particle wave functions have an $x$-independent normalization, one clearly has 
\begin{eqnarray}
D_N(x_1,
\dots,x_N)= \prod_i \int d^ 2 b_i 
|\Psi_{\vec n_{\perp}} (x_1,\vec b_1, \dots , x_N ,\vec b_N) |^2= 
|\varphi(x_1 , \dots , x_N)|^2 .
\end{eqnarray}
Note that here the variables $x_i$ are fixed, i.e., they are not
dynamical, and we can still multiply by an arbitrary function
$\varphi(x_1,\dots ,x_N)$ which would be generated after
requantization by some unspecified 1-dimensional dynamics, say $M^2 = M_{\rm HO}^2 + M_{1+1}^2$. Our model in Eq.~(\ref{eq:D3})
corresponds to the product $\varphi(x_1, \dots x_N)=\prod_i
\psi(x_i)$. This is similar to the Born-Oppenheimer approximation
where the longitudinal variables obbey classical dynamics. In this
extreme case the mass terms give after minimization $x_i = m_i/m$,
with $m=\sum_i m_i$ yielding a contribution $m^2$ to $M^2$ and $D_N(x_1,
\dots,x_N) \to \prod_i \delta(x_i - m_i/m) $, which for equal masses
corresponds to locating the distridution in our Eq.~(\ref{eq:D3}) at the maximum,
$x_i=1/N$.  That model breaks the transverse translational
invariance, which can be restored by projecting onto zero total
transverse momentum using the Peierls-Yoccoz
method~\cite{Betz:1983dy}, such that for $m_i=0$ one finds $ M_N^2 = 2 N \sigma $
for the symmetric state. A straightforward calculation with the rest
frame wave function yields the following result for the unintegrated dPDF:
\begin{eqnarray}
D_2 (x_1,x_2,\vec k_\perp) = \int d^2 b d^2 B |\Psi_0 (x_1,\vec B+ \frac12 \vec b , x_2,\vec B- \frac12 \vec b) |^2 e^{i \vec b \cdot \vec k_\perp}=  
D_2(x_1,x_2) e^{-\frac{\vec
    k_\perp^2}{4\sigma}\frac{x_1+x_2}{x_1 x_2}}.
\end{eqnarray}
Thus, there exists a transverse dynamics given by Eq.~(\ref{eq:M2})
where the effective 1-dimensional treatment used here holds, but it
does not support the often assumed transverse-longitudinal
factorization.

\bibliography{dPDF}

\end{document}